\documentclass[aps,prb,showpacs,twocolumn,floats]{revtex4}

\usepackage{graphicx,psfig}
\usepackage{dcolumn}
\usepackage{bm}

\begin{document}
\title{Effects of nonlinear sweep in the Landau-Zener-Stueckelberg effect}
\author{D. A. Garanin and R. Schilling}
 \affiliation{ Institut f\"ur Physik, Johannes-Gutenberg-Universit\"at,
 D-55099 Mainz, Germany}

\begin{abstract}
We study the Landau-Zener-Stueckelberg (LZS) effect for a
two-level system with a time-dependent nonlinear bias field (the
sweep function) $W(t)$. Our main concern is to investigate the
influence of the nonlinearity of $W(t)$ on the probability $P$ to
remain in the initial state. The dimensionless quantity
$\varepsilon =\pi \Delta ^{2}/(2\hbar v)$ depends on the coupling
$\Delta $ of both levels and on the sweep rate $v$. For fast sweep
rates, i.e., $\varepsilon \ll 1,$ and monotonic, analytic sweep
functions linearizable in the vicinity of the resonance we find
the transition probability $1-P\cong \varepsilon (1+a)$, where
$a>0$ is the correction to the LSZ result
 due to the nonlinearity
of the sweep. Further increase of the sweep rate with nonlinearity
fixed brings the system into the nonlinear-sweep regime
characterized by $1-P\cong \varepsilon ^{\gamma }$ with $\gamma
\neq 1,$ depending on the type of sweep function. In case of slow
sweep rates, i.e., $\varepsilon \gg 1$, an interesting
interference phenomenon occurs. For analytic $W(t)$ the
probability $P=P_{0}e^{-\eta }$ is determined by the singularities
of $\sqrt{\Delta ^{2}+W^{2}(t)}$ in the upper complex plane of
$t$. If $W(t)$ is close to linear, there is only one singularity,
that leads to the LZS result $P=e^{-\varepsilon }$ with important
corrections to the exponent due to nonlinearity. However, for,
e.g., $W(t)\propto t^{3}$ there is a pair of singularities in the
upper complex plane. Interference of their contributions leads to
oscillations of the prefactor $P_{0}$ that depends on the sweep
rate through $\varepsilon $ and turns to zero at some $\varepsilon
$. Measurements of the oscillation period and of the exponential
factor would allow to determine $\Delta ,$ independently.
\end{abstract}

 \maketitle

\section{\protect\bigskip Introduction}

\begin{figure}[t]
\unitlength1cm
\begin{picture}(11,6)
\centerline{\psfig{file=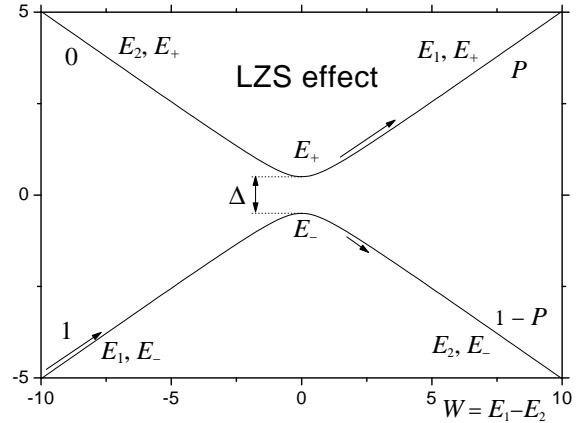,angle=-90,width=9cm}}
\end{picture}
\caption{ \label{LZEffect} A pair of tunnel-splitted levels vs.
energy bias $W(t)$. Here $E_1$ and $E_2$ are the bare energy
levels ($\Delta=0$), whereas $E_\pm$ are the exact adiabatic
energy levels of Eq.\ (\protect\ref{epspmDef}).
$P$ denotes the probability to remain in the (bare) state 1 after crossing the resonance.
}
\end{figure}

We will study the Landau-Zener-Stueckelberg (LZS) problem of
quantum-mechanical transitions between the levels of a two-level
system at the avoided level crossing (see Fig.\ \ref{LZEffect})
caused by the time dependence of the Hamiltonian
\cite{lan32,zen32,stu32,akusch92,dobzve97}
\begin{equation}\label{Ham}
\widehat{H}=-\frac{1}{2}W(t)\sigma _{z}+\frac{1}{2}\Delta \sigma _{x},
\end{equation}
where $\sigma _{\alpha },$ $\alpha =x,y,z$ are the Pauli matrices and
\begin{equation}
W(t)\equiv E_{1}(t)-E_{2}(t)  \label{WtDef}
\end{equation}
is the time-dependent bias of the two bare ($\Delta =0)$ energy
levels.  Because of its generality, the LZS problem is pertinent
to many areas of physics. In particular, the time-dependent model
of Eq.\ (\ref{Ham}) is a simplification of the two-channel
scattering problem with the curve crossing in the theory of
molecular collisions (see, e.g., Refs.\ \onlinecite{chi74,nak02}
and references therein). An appropriate choice of $W(t)$,
including oscillative time dependences,
\cite{groditjunhae91,kay93,kay94,miysairae98,ternak98} may allow
to manipulate the evolution of the system in a controlled
way.\cite{groditjunhae91,ternak98,garsch02} The most important
case is probably that of the linear energy sweep:
\begin{equation}
W(t)=vt,\qquad v={\rm const>0}  \label{LinearSweep}
\end{equation}
that can be solved exactly. If the system at $t\rightarrow -\infty $ was in
the bare ground state $\psi _{1}=|\downarrow \rangle $ before crossing the
resonance, the probability to stay in this state after crossing the
resonance at $t\rightarrow \infty $ \ is given by \cite{zen32,stu32,dobzve97}
\begin{equation}
P(\infty )\equiv P=e^{-\varepsilon },\qquad \varepsilon \equiv \frac{\pi
\Delta ^{2}}{2\hbar v}  \label{PLZ}
\end{equation}
{\em for all} $\varepsilon$.\cite{LZFoot}  Solution of the LZS
problem with linear sweep for a general initial condition (the
complete scattering matrix) and for multilevel systems can be
found in  Refs.\ \onlinecite{nak87,poksin00}.

Arguing that a typical smooth sweep function $W(t)$ crossing the
resonance only one time can be linearized close to the
resonance\cite{lanlif3,dobzve97,poksin00} and that the
$t$-dependence of $W(t)$ far enough from the resonance does not
matter, the LZS problem has been in most cases solved for linear
sweeps $W(t)=vt$ (see, however, Ref.\ \onlinecite{ternak97}).
Recently, however, there has been a great deal of interest to the
LZS effect in systems of interacting tunneling species (see, e.g.,
Ref.\ \onlinecite{hamraemiysai00}) initiated by experiments on
molecular magnets.\cite{weretal00epl} In Ref.\
\onlinecite{hamraemiysai00} the LZS effect has been investigated
for an interacting system of $ N $ particles with spin $S=1/2$.
Since in this model each spin interacts with all the others with
the same coupling constant, one can use a mean-field approach
which becomes exact for $N\rightarrow \infty $. Therefore in
general the effective field becomes nonlinear in time, due to the
nonlinear $t$-dependence of the field on a given spin acting from
other spins.

It is thus of principal interest to investigate the two-level
Landau-Zener-Stueckelberg problem for a nonlinear energy sweep.
First, for a weakly nonlinear sweep there will be corrections to
Eq.\ (\ref{PLZ}) that should be especially pronounced in the
slow-sweep region $\varepsilon \gg 1.$ Since an experimental
realization of a linear sweep can only be approximate it is
important to determine these corrections.  Second, essentially
nonlinear sweeps, including those with multiple crossing the
resonance, are becoming technically feasible for such objects as
magnetic molecules, where the Hamiltonian can be easily tuned by
the external magnetic field (see, e.g., Ref.\
\onlinecite{werses99}). These nonlinear energy sweeps could be
used to manipulate the system in a desired way, say to ensure
$P=0$ for the {\em finite} sweep rates.

In our recent Letter \cite{garsch02} we have solved the {\em inverse}
Landau-Zener-Stueckelberg problem, i.e., we have found such sweep functions $%
W(t)$ that ensure the required time-dependent probabilities $P(t)$
to stay in the initial
state$.$ Among the solutions are those corresponding to the time-symmetric $%
P(t)$ satisfying $P(-\infty )=1$ and $P(\infty )\equiv P=0.$ The functions $%
W(t)$ are more complicated than corresponding $P(t)$ and they become
cumbersome for more sophisticated forms of $P(t).$ An interesting task is to
find general criteria for the complete conversion $\psi _{1}\rightarrow \psi
_{2},$ i.e., for $P=0,$ for the direct LZS problem and to apply it to the
simplest nonlinear forms of $W(t).$ This is a further aim of our paper. We
will obtain analytical solutions for fast and slow sweep as well as
numerical solutions in the whole range of parameters.

The outline of the paper is as follows. In the next section we
will describe the LZS problem \ for a general sweep. Sec.\
\ref{sec-fastsweep} contains the discussion of the fast sweep,
whereas results for slow sweep are presented in Sec.\
\ref{sec-slowsweep}. We will show that in the slow-sweep case
different kinds of singularities that characterize $W(t)$
sensitively influence the propability $P.$ The final section
contains a summary and some conclusions.

For numerical calculation throughout the paper we use Wolfram
Mathematica 4.0 that employs a very accurate differential equation
solver needed for dealing with strongly oscillating solutions over
large time intervals.

\section{LZS problem for general sweep}

The Schr\"{o}dinger equation for the coefficients of the wave function $\psi
(t)=c_{1}(t)\psi _{1}+c_{2}(t)\psi _{2}$ can be written as
\begin{eqnarray}
i\hbar \frac{d}{dt}\tilde{c}_{1}(t) &=&\frac{\Delta }{2}\tilde{c}%
_{2}(t),  \nonumber \\
i\hbar \frac{d}{dt}\tilde{c}_{2}(t) &=&-W(t)\tilde{c}_{2}(t)+\frac{\Delta }{2%
}\tilde{c}_{1}(t),  \label{c1c2Eq}
\end{eqnarray}
where
$\tilde{c}_{1,2}(t)\equiv c_{1,2}(t)\exp \left[ -\frac{i}{2\hbar }%
\int_{-\infty }^{t}dt^{\prime }W(t^{\prime })\right] $ satisfy the
initial conditions
\begin{equation}\label{InitCond}
\tilde{c}_{1}(-\infty )=1, \qquad \tilde{c}_{2}(-\infty )=0,
\end{equation}
 and $W(t)\equiv E_{1}(t)-E_{2}(t)$ satisfies
\mbox{$W(-\infty )=-\infty $.} A general nonlinear sweep function
$W(t)$ can be parametrized as follows
\begin{equation}
W(t)=\Delta w(u),\qquad u\equiv vt/\Delta .  \label{WbarDef}
\end{equation}
The corresponding dimensionless form of Eq.\ (\ref{c1c2Eq}) is
\begin{eqnarray}
\tilde{c}_{1}^{\prime }(u) &=&-\frac{i\tilde{\varepsilon}}{2}\tilde{c}%
_{2}(u),  \nonumber \\
\tilde{c}_{2}^{\prime }(u) &=&i\tilde{\varepsilon}w(u)\tilde{c}_{2}(u)-\frac{%
i\tilde{\varepsilon}}{2}\tilde{c}_{1}(u)  \label{c1c2Equ}
\end{eqnarray}
where $^{\prime }$ denotes differentiation with respect to $u$ and the
sweep-rate parameter $\tilde{\varepsilon}$ is defined by
\begin{equation}
\tilde{\varepsilon}\equiv \frac{\Delta ^{2}}{\hbar v}=\frac{2}{\pi }%
\varepsilon .  \label{epstildeDef}
\end{equation}
(The reader should not confuse $\tilde{\varepsilon}$ and $\varepsilon $ that
differ by a numerical factor; Whereas $\tilde{\varepsilon}$ arises in a
natural way, $\varepsilon $ is used to represent most of the final results.)

It is convenient to represent dimensionless analytical functions $w(u)$ that
behave linearly near $u=0$ in the form
\begin{equation}
w(u)=\alpha ^{-1}f(\alpha u)\cong u+\frac{\alpha }{2!}f_{0}^{\prime \prime
}u^{2}+\frac{\alpha ^{2}}{3!}f_{0}^{\prime \prime \prime }u^{3}+\ldots
\label{fDef}
\end{equation}
where $f_{0}^{\prime }=1$ and, in general, $f_{0}^{\prime \prime
}\sim f_{0}^{\prime \prime \prime }\sim 1$. Whereas $v$ in Eq.\
(\ref{WbarDef}) stands for the sweep rate, the parameter $\alpha $
in (\ref{fDef}) controls the nonlinearity of the sweep. Below we
will work out explicit results for the{\em \ analytical} sweep
function
\begin{equation}
w(u)=\alpha ^{-1}\sinh (\alpha u)  \label{Wbarsinh}
\end{equation}
that satisfies $w(-u)=-w(u)$. A more general {\em \ analytical }sweep
function is
\begin{equation}
w(u)=\frac{1}{\alpha +\beta }\left( e^{\alpha u}-e^{-\beta u}\right)
\label{Sweepalphabeta}
\end{equation}
that is characterized by $w(-u)\neq -w(u)$ for $\alpha \neq \beta .$ An
interesting property of the corresponding LZS problem is that for $\alpha
\neq \beta $ interchanging $\alpha $ and $\beta $ leads to essentially
different solutions of the corresponding Schr\"{o}dinger equation while $P$
being the same in both cases. This can be proven by considering the general
scattering matrix of the problem, similarly to the proof of the reflection
and transmission coefficients for scattering on a nonsymmetric
one-dimensional potential being the same for both directions of the ongoing
particle.\cite{lanlif3}

Also we would like to consider nonanalytical sweep functions like the
power-law $w(u)=u^{\alpha }$ with $\alpha >0.$ \ It is convenient to put
this problem into a more general form and to introduce {\em crossing} and
{\em returning} sweeps
\begin{equation}
w(u)=\left\{
\begin{array}{ll}
{\rm sign}(u)|u|^{\alpha }, & \text{crossing} \\
-|u|^{\alpha }, & \text{returning.}
\end{array}
\right.  \label{WbarPowDef}
\end{equation}
Note that returning sweeps (with double crossing the resonance) are
naturally realized in atomic and molecular collisions where the distance
between the colliding species at first decreases and than increases again.
\cite{stu32}

\section{Fast sweep}

\label{sec-fastsweep}

\begin{figure}[t]
\unitlength1cm
\begin{picture}(11,6)
\centerline{\psfig{file=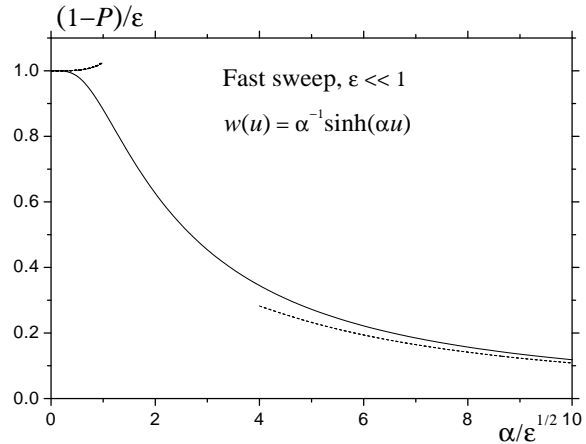,angle=-90,width=9cm}}
\end{picture}
\caption{ \label{FastSweepCrossover} Dependence of the transition
probability $1-P$ on the nonlinearity parameter $\alpha$ for the
sweep function $w(u)=\alpha^{-1}\sinh(\alpha u)$ in the fast-sweep
regime, $\varepsilon\ll 1$. The dashed lines are the asymptotes of
Eqs.\   (\protect\ref{PsmallepsNonlinear}) and
(\protect\ref{Psmallepsilonsinh}). }
\end{figure}

In this section we will discuss the behavior of $P$ for fast sweep
for which the propability $P$ should stay close to 1. The form of
Eq.\ (\ref{c1c2Equ})
is convenient to perform the fast-sweep approximation $\tilde{\varepsilon}%
\ll 1.$ In zeroth order of the perturbation theory one has $\tilde{c}%
_{1}(u)=1$ which can be used to obtain $\tilde{c}_{2}(u)$ from the
second line of Eq.\ (\ref{c1c2Equ}). The resulting equation for
$\tilde{c}_{2}(u)$ can easily be solved. Using this result, the
probability to stay in the initial state 1 can be expressed as
\begin{eqnarray}
&&P \cong 1-|\tilde{c}_{2}(\infty )|^{2} \nonumber \\
&& =1  -\frac{\tilde{\varepsilon}^{2}}{4}\left| \int_{-\infty
}^{\infty
}du\exp \left[ -i\tilde{\varepsilon}\int_{0}^{u}du^{\prime }w(u^{\prime })%
\right] \right| ^{2}.  \label{PFastSweep}
\end{eqnarray}
One can see that this is not a standard perturbation theory that
would yield a correction of order $\tilde{\varepsilon}^{2}$ to
$P.$ The integral over $u$ in Eq.\ (\ref{PFastSweep}) assumes
large values for $\tilde{\varepsilon}\ll 1$ and is in general
non-analytic in $\tilde{\varepsilon}.$ Since $u\sim u_\varepsilon
\sim \varepsilon^{-1/2} \gg 1$ contribute to this integral for
$\tilde{\varepsilon}\ll 1,$ the latter is very sensitive to the
nonlinearity of $w(u).$ For weakly-nonlinear $w(u)$ that can be
expanded into the Taylor series of Eq.\ (\ref{fDef}) one has
\[
\int_{0}^{u}duw(u)\cong \frac{1}{2!}u^{2}+\frac{\alpha }{3!}f_{0}^{\prime
\prime }u^{3}+\frac{\alpha ^{2}}{4!}f_{0}^{\prime \prime \prime
}u^{4}+\ldots
\]
If $\alpha \ll u_{\varepsilon }^{-1}\sim
\tilde{\varepsilon}^{1/2},$ it is the first term of this expansion
that is dominating. One can transform
this condition into the dimensional form if one writes $W(t)\cong \dot{W}%
_{0}t+\ddot{W}_{0}t^{2}/2!+\ldots $ and makes comparison with Eq.\ (\ref{fDef}%
). This yields the weak-nonlinearity condition
\begin{equation}
\alpha \ll \tilde{\varepsilon}^{1/2}\qquad \Longleftrightarrow \qquad \hbar
\ddot{W}_{0}^{2}\ll \dot{W}_{0}^{3}.  \label{LinCondPhys}
\end{equation}
With the help of Eqs.\ (\ref{WbarDef}) and (\ref{epstildeDef}) one
can
establish that the LZS transition takes place {\em not} in the range $%
W(t)\cong vt\sim \Delta $ around the resonance, as could be expected, but in
the much wider range
\begin{equation}
W(t)\sim \Delta /\sqrt{\tilde{\varepsilon}}=\sqrt{\hbar v}\gg \Delta
\label{ResRangeFastSweep}
\end{equation}
for the fast sweep. In the weakly nonlinear regime the main
contribution to the integral in Eq.\ (\ref{PFastSweep}) stems from
$u\sim u_{\varepsilon }$ and one obtains
\begin{equation}
P\cong 1-\varepsilon \left\{ 1+\left[ \frac{\pi }{32}\left( f_{0}^{\prime
\prime \prime }-\frac{5}{3}f_{0}^{\prime \prime 2}\right) \right] ^{2}\left(
\frac{\alpha ^{2}}{\varepsilon }\right) ^{2}\right\} .
\label{PsmallepsNonlinear}
\end{equation}
Since $\alpha ^{2}\ll \varepsilon ,$ the leading-order result for
$P$ does not depend on the nonlinearity, as expected. But Eq.\
(\ref {PsmallepsNonlinear}) also demonstrates that the
next-to-leading order in the nonlinearity {\em reduces} the
probability $P,$ {\em independently} on whether the sweep function
$w(u)$ [or $W(t)$] grows slower or faster than
linear, in contrast to the expectation. This effect that dependes on $%
f_{0}^{\prime \prime }$ and $f_{0}^{\prime \prime \prime }$ can be
rather small, though (see Fig.\ \ref{FastSweepCrossover}).

If $\alpha \gg \varepsilon ^{1/2}$ and  $w(u)$ is a power series
that terminates at finite order $u^{n},$ then it is this latter
term that
dominates, and the main contribution to the integral in Eq.\ (\ref{PFastSweep}%
) comes from $u_{\varepsilon }\sim {\varepsilon}^{-1/(n+1)}$ that
yields
\begin{equation}
1-P\sim \left( \varepsilon /\alpha ^{2}\right) ^{2n/(n+1)}.
\label{Psmallepsilonn}
\end{equation}
The range of the energy bias $W(t)$ that is responsible for
tunneling is given by
\begin{equation}
W(t)\sim \Delta /{\varepsilon}^{n/(n+1)}\gg \Delta
\label{rangeofbias}
\end{equation}
[cf. Eq.\ (\ref{ResRangeFastSweep})].

A particular case of the above is the {\em crossing} power-law
sweep described by Eq.\ (\ref{WbarPowDef})
that contains the linear sweep as a particular case. Here one has $%
\int_{0}^{u}duw(u)=|u|^{\alpha +1}/(\alpha +1)$ \ and Eq.\
(\ref{PFastSweep}) yields
\begin{equation}
P\cong 1-\Gamma ^{2}\left( \frac{1}{1+\alpha }\right) \left( \frac{\tilde{%
\varepsilon}}{1+\alpha }\right) ^{2\alpha /(1+\alpha )},
\label{PPowCrossing}
\end{equation}
where $\Gamma (x)$ is the gamma function. In the case $\alpha =1$ the
familiar expansion $P\cong 1-\varepsilon $ of Eq.\ (\ref{PLZ}) for $%
\varepsilon \ll 1$ is recovered. For $\alpha >1$  the transition
probability $1-P$ is smaller than that for the linear sweep. The
opposite result is obtained for $\alpha <1,$ e.g.,
$1-P\propto \sqrt{\tilde{\varepsilon}}$ for $\alpha =1/3.$ For the {\em %
returning} power-law sweep of Eq.\ (\ref{WbarPowDef}) one has $%
\int_{0}^{u}duw(u)=-{\rm sign}(u)|u|^{\alpha +1}/(\alpha +1)$ and the result
has the form
\begin{equation}
P\cong 1-\Gamma ^{2}\left( \frac{1}{1+\alpha }\right) \cos ^{2}\left( \frac{%
\pi }{2(1+\alpha )}\right) \left( \frac{\tilde{\varepsilon}}{1+\alpha }%
\right) ^{2\alpha /(1+\alpha )}.  \label{PPowReturning}
\end{equation}

For the sweep described by Eq.\ (\ref{Wbarsinh}) one has $%
\int_{0}^{u}duw(u)=(\cosh \alpha u-1)/\alpha ^{2}.$ In the case
$\alpha ^{2}\ll \tilde{\varepsilon}$ Eq.\ (\ref{PFastSweep})
yields the well known result for the linear sweep $P\cong
1-\varepsilon ,$ as it should be for small enough nonlinearity. In
the opposite case one can use the approximation
$\int_{0}^{u}du'w(u')\cong e^{\alpha u}/(2\alpha ^{2})$ to obtain
\begin{equation}
P\cong 1-\frac{\tilde{\varepsilon}^{2}}{\alpha ^{2}}\left( \ln \frac{2\alpha
^{2}}{\tilde{\varepsilon}}-\gamma \right) ^{2},\qquad \tilde{\varepsilon}\ll
\alpha ^{2},  \label{Psmallepsilonsinh}
\end{equation}
where $\gamma =0.577216.$ This is in accord with Eq.\
(\ref{Psmallepsilonn}) in the limit $n\rightarrow \infty ,$ up to
the logarithm. The crossover between the linear and nonlinear
regimes for the sweep function of Eq.\ (\ref {Wbarsinh}) on
$\alpha /\sqrt{\varepsilon }$ is shown in Fig.\
\ref{FastSweepCrossover}.

\section{Slow sweep}

\label{sec-slowsweep}

In this section we will study the asymptotic behavior of $P$ for $%
\varepsilon \gg 1,$ i.e., for slow sweeps. This behavior will strongly
depend on the analytical properties of $w(u).$ Let us first describe the
general situation in case of slow sweep.

\subsection{General}

It is convenient to solve the Schr\"{o}dinger equation in the
adiabatic basis formed of the states $\psi _{\pm }(t)$ that are
solutions of the stationary Schr\"{o}dinger equation at the time
$t$. The evolution of the system nearly follows the lowest, i.e.,
$E_{-}(t),$ of the adiabatic energy levels (see Fig.\
\ref{LZEffect})
\begin{equation}\label{epspmDef}
E_{\pm }(t) =\pm\frac{1}{2} \Omega (t) =\pm\frac{1}{2}
\sqrt{W^{2}(t)+\Delta ^{2}},
\end{equation}
whereas the probability of the transition to the upper level $E_{+}$ is
small. The explicit form of the adiabatic states is \cite{garsch02}
\[
\psi _{\pm }(t)=\frac{1}{\sqrt{2}}\left[ \pm K_{\pm }(t)\psi _{1}+K_{\mp
}(t)\psi _{2}\right] ,
\]
where $K_{\pm }(t)\equiv \sqrt{1\pm W(t)/\Omega (t)}.$ Writing the
wave function in the form $\psi =c_{+}(t)\psi _{+}(t)+c_{-}(t)\psi
_{-}(t)$ one obtains the equations for $\tilde{c}_{\pm }(t)\equiv
\exp \left( i\int dtE_{-}(t)/\hbar \right) c_{\pm }(t).$ The
dimensionless form of these equations reads [see Eq.\
(\ref{WbarDef})]
\begin{eqnarray}
&&\tilde{c}_{-}^{\prime }(u) =\frac{w^{\prime }(u)}{2\overline{\Omega }%
^{2}(u)}\tilde{c}_{+}(u),\qquad \overline{\Omega }(u)\equiv \sqrt{1+w^{2}(u)}
\nonumber \\
&&\tilde{c}_{+}^{\prime }(u) =-i\tilde{\varepsilon}\overline{\Omega }(u)%
\tilde{c}_{+}(u)-\frac{w^{\prime }(u)}{2\overline{\Omega }^{2}(u)}\tilde{c}%
_{-}(u)  \label{cpcmEqsu}
\end{eqnarray}
and the initial condition is $\tilde{c}_{-}(-\infty )=1,$ $\tilde{c}%
_{+}(-\infty )=0.$ The probability $\ P$ to stay in state 1 is given by
\begin{equation}
P=\left\{
\begin{array}{ll}
|\tilde{c}_{+}(\infty )|^{2}, & \text{crossing} \\
|\tilde{c}_{-}(\infty )|^{2}=1-|\tilde{c}_{+}(\infty )|^{2}, & \text{%
returning}
\end{array}
\right.   \label{PAdiDef}
\end{equation}
and it is small for crossing sweeps and close to 1 for returning sweeps.
In fact, for returning sweeps $P$ tends to 1 in the limits of both
fast and slow sweeps, as is illustrated in Fig.\
\ref{Ppowalsmall}.
Indeed, from Fig.\ \ref{LZEffect} one can see that for a fast
sweep the system practically remains on the bare level $E_1$,
whereas for a slow sweep it travels along the adiabatic level
$E_-$ and thus returns to $E_1$ for $W(\infty)=-\infty$.

One possible way of solving Eq.\ (\ref{WbarDef}) is to transform
it to a single second-order differential equation and then to
apply the WKB approximation to the corresponding
overbarrier-reflection problem with a complex potential. The
result is a linear combination of two solutions that can be
interpreted as ongoing and reflected waves. Then the
(exponentially small) factor in front of the reflected-wave
solution can be found from the analysis of Stokes lines in the
complex plane (see, e.g., Ref.\ \onlinecite{nak02} and references
therein). This analysis is rather involved, however. Below we will
present an alternative and more simple method of solving Eq.\
(\ref{WbarDef}) that does not rely on the WKB approximation.

One can immediately write down the formal solution of  Eq.\
(\ref{WbarDef}) for $|\tilde{c}_{+}(\infty )|^{2}$ that contains
yet to be determined $\tilde{c}_{-}(u )$
\begin{equation}
|\tilde{c}_{+}(\infty )|^{2}= \left| \frac{1}{2}\int_{-\infty
}^{\infty }du\frac{w^{\prime }(u)}{\overline{\Omega
}^{2}(u)}\tilde{c}_{-}(u)\exp \left[ i\tilde{\varepsilon}\Phi
(u)\right] \right| ^{2}  \label{ctilpgen}
\end{equation}
with
\begin{equation}
\Phi (u)\equiv \int_{0}^{u}du'\,\overline{\Omega }(u').
\label{SuDef}
\end{equation}
Since for the slow sweep $\tilde{c}_{-}(u)$ remains close to 1, $\tilde{c}%
_{-}(u)\Rightarrow 1$ is a reasonable approximation.
We will see below that this approximation is sufficient to obtain
the correct exponential in the exponentially small $P$, as well as
the prefactor with better than 10\% accuracy.
On the top of it, the approximation will be refined to obtain the
exact prefactor.
The asymptotic $\tilde{%
\varepsilon}$ dependence of the rhs of Eq.\ (\ref{ctilpgen})
depends strongly on the analytical properties of $\Phi (u).$ Even
if $w(u)$ is chosen to be an analytical function, the integrand
$\overline{\Omega }(u)$ in Eq.\ (\ref {SuDef}) is not analytical
but has singularities at the branch points at which $w^{2}(u)=-1.$
We will see that these singularities determine the
large-$\varepsilon $ dependence of $P,$ for analytical sweep functions $%
w(u).$

\subsection{Non-analytic sweep functions}
\label{sec-nonanalyticslow}

\begin{figure}[t]
\unitlength1cm
\begin{picture}(11,6)
\centerline{\psfig{file=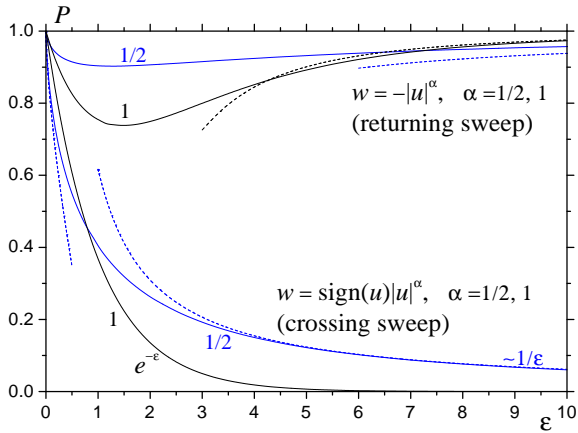,angle=-90,width=9cm}}
\end{picture}
\caption{ \label{Ppowalsmall} Staying probability $P(\varepsilon)$
for power-law sweep functions with exponents $\alpha=1/2$ and
$\alpha=1$. Solid lines are numerical results and dashed lines are
some of the analytical asymptotes. }
\end{figure}

\begin{figure}[t]
\unitlength1cm
\begin{picture}(11,6)
\centerline{\psfig{file=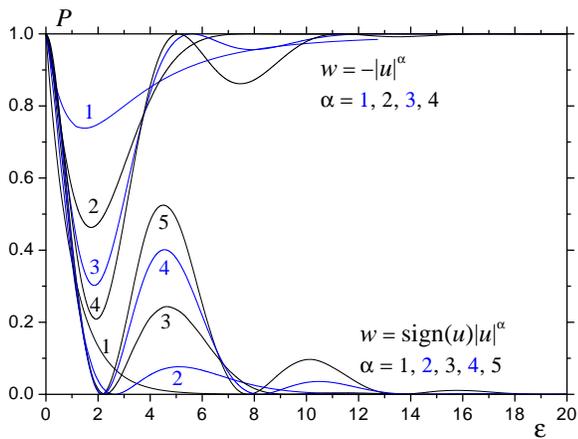,angle=-90,width=9cm}}
\end{picture}
\caption{ \label{Ppowallarge} $P(\varepsilon)$ for power-law sweep
functions with exponents $\alpha=1,2,3,4$ and $5$. }
\end{figure}

\begin{figure}[t]
\unitlength1cm
\begin{picture}(11,6)
\centerline{\psfig{file=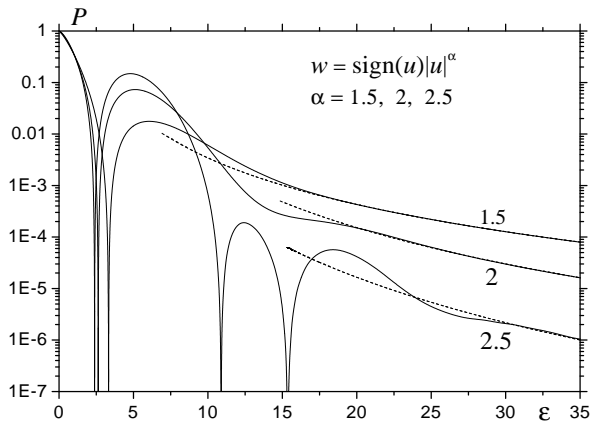,angle=-90,width=9cm}}
\end{picture}
\caption{ \label{PpowallargeLog} $P(\varepsilon)$ for crossing
power-law sweep functions with exponents $\alpha=1.5,2$ and $2.5$.
Dashed lines are high-$\varepsilon$ asymptotes of Eq.\
(\protect\ref{PAdiSingRes}). }
\end{figure}

We consider for the beginning the simpler case of sweep functions that are
non-analytic at cros$\sin $g the resonance, $u=0,$ namely the power-law
sweep functions of Eq.\ (\ref{WbarPowDef}) with a general $\alpha .$ For $%
\tilde{\varepsilon}\gg 1$ the integral in Eq.\ (\ref{ctilpgen}) with $\tilde{c%
}_{-}(u)\Rightarrow 1$ is dominated by small $u$ for which
$w(u)\ll 1,$ and thus $\overline{\Omega }(u)\cong 1.$ After taking
advantage of the symmetry of Eq.\ (\ref{WbarPowDef}) it simplifies
to
\[
|\tilde{c}_{+}(\infty )|^{2}=%
\mathrel{\mathop{\lim }\limits_{\delta \rightarrow 0}}%
\left| \int_{0}^{\infty }du\,e^{-\delta u}\alpha u^{\alpha -1}\left\{
\begin{array}{l}
\cos \left( \tilde{\varepsilon}u\right) \\
\sin (\tilde{\varepsilon}u)
\end{array}
\right\} \right| ^{2}
\]
for crossing and returning sweeps, respectively. The final result
for $P$ of Eq.\ (\ref{PAdiDef}) reads
\begin{equation}
P\cong \left\{ \renewcommand{\arraystretch}{2.5}
\begin{array}{ll}
\displaystyle\frac{\Gamma ^{2}(1+\alpha )}{\tilde{\varepsilon}^{2\alpha }}%
\cos ^{2}\left( \frac{\pi \alpha }{2}\right) , & \text{crossing} \\
\displaystyle1-\frac{\Gamma ^{2}(1+\alpha )}{\tilde{\varepsilon}^{2\alpha }}%
\sin ^{2}\left( \frac{\pi \alpha }{2}\right) , & \text{returning.}
\end{array}
\right.  \label{PAdiSingRes}
\end{equation}
The nonanalytic $\varepsilon $ dependent contribution here stems
from the nonanalytic behavior of $w(u)$ at $u=0$ and it vanishes
for analytic
functions such as $w(u)=u$ or $w(u)=u^{3}$ for the crossing sweep and $%
w(u)=u^{2}$ for the returning sweep. Numerical results for $P(\varepsilon )$
along with the asymptotes of Eqs.\ (\ref{PPowCrossing}), (\ref{PPowReturning}%
), and (\ref{PAdiSingRes}) are shown for $\alpha =1/2$ and $\alpha =1$ in
Fig.\ \ref{Ppowalsmall}. One can see that the slow-sweep asymptote of Eq.\ (%
\ref{PAdiSingRes}) works well starting from relatively low values of $%
\varepsilon $ for crossing sweeps and from larger $\varepsilon $
for returning sweeps. $P(\varepsilon )$ for $\alpha =1,2,3,4$ and
$5$ are shown in Fig.\ \ref{Ppowallarge}. In this case the
asymptotes of Eq.\ (\ref {PAdiSingRes}) require rather large
values of $\varepsilon $ and they can be
seen on the log plot only (see Fig.\ \ref{PpowallargeLog}) With increasing $%
\alpha $ oscillations of $P(\varepsilon )$ develop; the reason for this will
be explained below.

\subsection{Analytic sweep functions}

As we will see, the form of the probability $P$ for $\varepsilon \gg 1$
crucially depends on the number of singularities of $\overline{\Omega }(u)=%
\sqrt{1+w^{2}(u)}$ in the upper complex plane. In the next
subsection we will discuss sweep functions $w(u)$ for which there
is only one singularity, starting from the weakly nonlinear case.
For strongly nonlinear sweep with retardation in the resonance
region, there is a pair of singularities symmetric with respect to
the imaginary axis, that dominate $P$ for $\varepsilon \gg 1$. In
this case $P$ can be represented in the form of a small
exponential with a prefactor oscillating as a function of
$\varepsilon $.

\subsubsection{One singularity: Exponent}

For analytic sweep functions $w(u)$ the value of $|\tilde{c}_{+}(\infty
)|^{2}$ given by Eq.\ (\ref{ctilpgen}) is exponentially small for $\tilde{%
\varepsilon}\gg 1,$ and it is defined by the singularities of the integrand
in the upper complex half-plane that are closest to the real axis. These
singularities are combinations of poles and branching points at $%
w^{2}(u)=w_{c}^{2}=-1.$ Let us consider at first sweep functions
$w(u)$ that are monotonic and close to the linear function, $w(u)$
$\approx u.$ In this case there is only one singularity at
$u\approx w(u)=w_{c}=i,$ and it is convenient to use $w$ as the
sweep variable instead of $u$ in Eq.\ (\ref {ctilpgen}):
\begin{eqnarray}
&&P\cong \left| \frac{1}{2}\int_{-\infty }^{\infty }\frac{dw}{\overline{%
\Omega }^{2}(w)}\tilde{c}_{-}(w)e^{i\tilde{\varepsilon}\Phi (w)}\right| ^{2}
\nonumber \\
&&\Phi (w)\equiv \int_{0}^{w}dw\frac{du(w)}{dw}\overline{\Omega }(w).
\label{ctilpgenW}
\end{eqnarray}
We deform the integration line into the contour ${\cal C}$ that goes down
along the left side of the cut $[i,i\infty ],$ then around the pole at $w=i,$
and finally along the right side of the cut to $i\infty .$ Then the phase $%
\Phi (w)$ can be written in the form
\begin{eqnarray}
&&\Phi (w)=\Phi _{c}+\delta \Phi (w)  \nonumber \\
&&\Phi _{c}=\int_{0}^{i}dw\frac{du(w)}{dw}\overline{\Omega }%
(w)=i\int_{0}^{1}dy\,u^{\prime }(y)\sqrt{1-y^{2}}  \nonumber \\
&&\delta \Phi (y)=i\int_{1}^{y}dy\,u^{\prime }(y)\sqrt{1-y^{2}},\qquad w=iy.
\label{SWDef}
\end{eqnarray}
Accordingly the result of integration in Eq.\ (\ref{ctilpgenW})
can be represented in the form
\begin{equation}
P=P_{0}e^{-\eta },\qquad \eta =2\tilde{\varepsilon}%
\mathop{\rm Im}%
\Phi _{c},  \label{PP0Def}
\end{equation}
where the prefactor $P_{0}$ follows from
\begin{equation}
P_{0}=\left| \frac{1}{2}\int_{{\cal C}}\frac{dy}{1-y^{2}}\tilde{c}_{-}(y)e^{i%
\tilde{\varepsilon}\delta \Phi (y)}\right| ^{2}.  \label{P0Def}
\end{equation}
Large values of $\tilde{\varepsilon}$ in our case ensure
exponential smallness of $P$ in Eq.\ (\ref{PP0Def}). Let us at
first calculate the exponent $\eta $ for a weakly-nonlinear sweep
described by Eq.\ (\ref{fDef}). Inverting Eq.\ (\ref{fDef}) for
$\alpha \ll 1$ one obtains
\[
u^{\prime }(y)\cong 1-i\alpha f_{0}^{\prime \prime }y+\frac{\alpha ^{2}}{2}%
\left[ f_{0}^{\prime \prime \prime }-3\left( f_{0}^{\prime \prime }\right)
^{2}\right] y^{2}
\]
and accordingly
\begin{equation}
\eta =2\tilde{\varepsilon}%
\mathop{\rm Im}%
\Phi _{c}\cong \varepsilon \left\{ 1+\frac{\alpha ^{2}}{8}\left[
f_{0}^{\prime \prime \prime }-3\left( f_{0}^{\prime \prime }\right) ^{2}%
\right] \right\} .  \label{etaquasilinear}
\end{equation}
For the time-antisymmetric sweep, $W(-t)=-W(t),$ one has
$f_{0}^{\prime \prime }=0.$ Then for $f_{0}^{\prime \prime \prime
}>0$ the exponent $\eta $ increases and the probability $P$ to
stay in state\ 1 decreases. This result seems counterintuitive
since in this case the system spends less time in the vicinity of
the resonance than in the case of a linear sweep and one could
expect that $P$ should increase. In fact, however, for a slow
sweep the (small) probability to occupy state (+) oscillates many
times during the resonance crossing and thus decreasing of $P$ is
a coherence effect that cannot be explaned by simple kinetic
arguments.

\subsubsection{One singularity: Prefactor}
\label{sec_onesingprefactor}

Now we turn to the calculation of the prefactor $P_{0}$ in Eq.\ (\ref{PP0Def}%
) setting $\tilde{c}_{-}(y)\Rightarrow 1$ in Eq.\ (\ref{P0Def})
[see comment below Eq.\ (\ref{SuDef})]. For
$\tilde{\varepsilon}\gg 1$ the values of $y$
in the vicinity of $1$ dominate the integrals. Thus one can introduce $%
t=y-1\ll 1$ and simplify Eq.\ (\ref{P0Def}) to
\begin{equation}
P_{0}\cong \left| \frac{1}{4}\int_{{\cal C}_{t}}\frac{dt}{t}\exp \left[
\tilde{\varepsilon}u_{c}^{\prime }\sqrt{2}\frac{2}{3}(-t)^{3/2}\right]
\right| ^{2},  \label{cplinft}
\end{equation}
where $u_{c}^{\prime }\equiv u^{\prime }(1)$ and the contour
${\cal C}_{t}$ goes from $\infty $ to 0 above the cut $[0,\infty
]$ around the pole at $t=0$ and then to $\infty $ below the cut.
At the upper and lower sides of the cut one has $(-t)_{\pm
}^{3/2}=\pm it^{3/2}.$ Thus Eq.\ (\ref{cplinft}) yields
\begin{eqnarray}
P_{0} &\cong &\left| \frac{1}{4}\left[ 2\pi i-2i\int_{0}^{\infty }\frac{dt}{t%
}\sin \left( \tilde{\varepsilon}u_{c}^{\prime }\sqrt{2}\frac{2}{3}%
t^{3/2}\right) \right] \right| ^{2}  \nonumber \\
&=&\left( \frac{\pi }{2}-\frac{\pi }{6}\right) ^{2}=\left( \frac{\pi }{3}%
\right) ^{2}\simeq 1.096.  \label{PAdiLin}
\end{eqnarray}
Note that this prefactor is a number and it does not depend on the argument
of sin, i.e., on $\tilde{\varepsilon}$ and on the constant $u_{c}^{\prime }$
that encapsulates the nonlinearity of the sweep. One can notice that this
prefactor {\em slightly} deviates from the exact prefactor 1 for the linear
sweep. \cite{zen32,dobzve97} This is the consequence of the approximation $%
\tilde{c}_{-}(w)\Rightarrow 1$ in Eq.\ (\ref{ctilpgenW}).

The reason for that wrong prefactor of Eq.\ (\ref{PAdiLin}) is the
following. Although $\tilde{c}_{-}(u)$ only slightly deviates from
1 along the real axis, it becomes singular at the relevant point
$w=i$ in the complex plane. This singularity can be worked out and
the prefactor can be corrected. To
this end, we note that the solution of Eq.\ (\ref{cpcmEqsu}) for $\tilde{%
\varepsilon}\gg 1$ can be expanded in powers of
$1/\tilde{\varepsilon}.$ This expansion captures only the
analytical part of the solution that yields $P=0$ in all orders in
$1/\tilde{\varepsilon}$ while the nonanalytic part of the solution
that yields $P\sim e^{-\varepsilon }$ cannot be found by this
method. Nevertheless, the $1/\tilde{\varepsilon}$ expansion is
sufficient to determine $\tilde{c}_{-}(w)$ that enters Eq.\
(\ref{P0Def}) to correct the prefactor. So we write down the
expansions for $\tilde{c}_{\pm }(w)$ in the form
\begin{equation}
\tilde{c}_{\pm }(w)=\sum_{n=0}^{\infty }\frac{\tilde{c}_{\pm ,n}(w)}{\left( i%
\tilde{\varepsilon}\right) ^{n}},  \label{cpmepsExp}
\end{equation}
where $\tilde{c}_{-,n}(-\infty )=\delta _{n0},$ and
$\tilde{c}_{+,n}(-\infty )=0.$ From Eq.\ (\ref{cpcmEqsu}) that can
be rewritten in the form
\begin{eqnarray}
\frac{d\tilde{c}_{-}(w)}{dw} &=&\frac{\tilde{c}_{+}(w)}{2\overline{\Omega }%
^{2}(w)},  \nonumber \\
\frac{d\tilde{c}_{+}(w)}{dw} &=&-i\tilde{\varepsilon}\frac{du(w)}{dw}%
\overline{\Omega }(w)\tilde{c}_{+}(w)-\frac{\tilde{c}_{-}(w)}{2\overline{%
\Omega }^{2}(w)}  \label{cpcmEqsW}
\end{eqnarray}
follows the infinite set of equations ($u^{\prime }\equiv du(w)/dw$)
\begin{eqnarray}
\frac{d\tilde{c}_{-,n}(w)}{dw} &=&\frac{\tilde{c}_{+,n}(w)}{2\overline{%
\Omega }^{2}}  \nonumber \\
\tilde{c}_{+,n}(w) &=&-\frac{1}{u^{\prime }\overline{\Omega }}\left[ \frac{d%
\tilde{c}_{+,n-1}(w)}{dw}+\frac{\tilde{c}_{-,n-1}(w)}{2\overline{\Omega }^{2}%
}\right]  \label{cpcmEqsWRec}
\end{eqnarray}
This set of equations can be solved recurrently:
\begin{eqnarray}
\tilde{c}_{+,0}(w) &=&0,\qquad \qquad \tilde{c}_{-,0}(w)=1,  \nonumber \\
\tilde{c}_{+,1}(w) &=&-\frac{1}{2u^{\prime }\overline{\Omega }^{3}},\qquad
\tilde{c}_{-,1}=-\frac{1}{4}\int_{-\infty }^{w}\frac{dw^{\prime }}{u^{\prime
}\overline{\Omega }^{5}},  \label{Recurrence}
\end{eqnarray}
etc. In Figs.\ \ref{fig-cpw}a and \ref{fig-cpw}b we compare the dependence $|%
\tilde{c}_{+}(w)|^{2}$ obtained by the numerical solution of Eq.\
(\ref
{cpcmEqsW}) and that using the $n=1$ term of Eq.\ (\ref{cpmepsExp}) with $%
\tilde{c}_{+,1}$ from Eq.\ (\ref{Recurrence}) \ for the linear sweep [$w(u)=u$%
] with $\varepsilon =15$. Whereas the agreement in the region $w\approx 0$
can be improved by taking into account further terms of Eq.\ (\ref{cpmepsExp}%
), the corresponding smooth and even functions $|\tilde{c}_{+}(w)|_{{\rm pert%
}}^{2}$ do not yield the asymptotic value $P=|\tilde{c}_{+}(\infty
)|^{2}$ at any order of $1/\tilde{\varepsilon}.$ The latter is due
to the singular part of the solution that oscillates and tends to
the exponentially small value $P=e^{-\varepsilon }$ [see Fig.\
\ref{fig-cpw}b]$.$

{\begin{figure}[t]
\unitlength1cm
\begin{picture}(11,6)
\centerline{\psfig{file=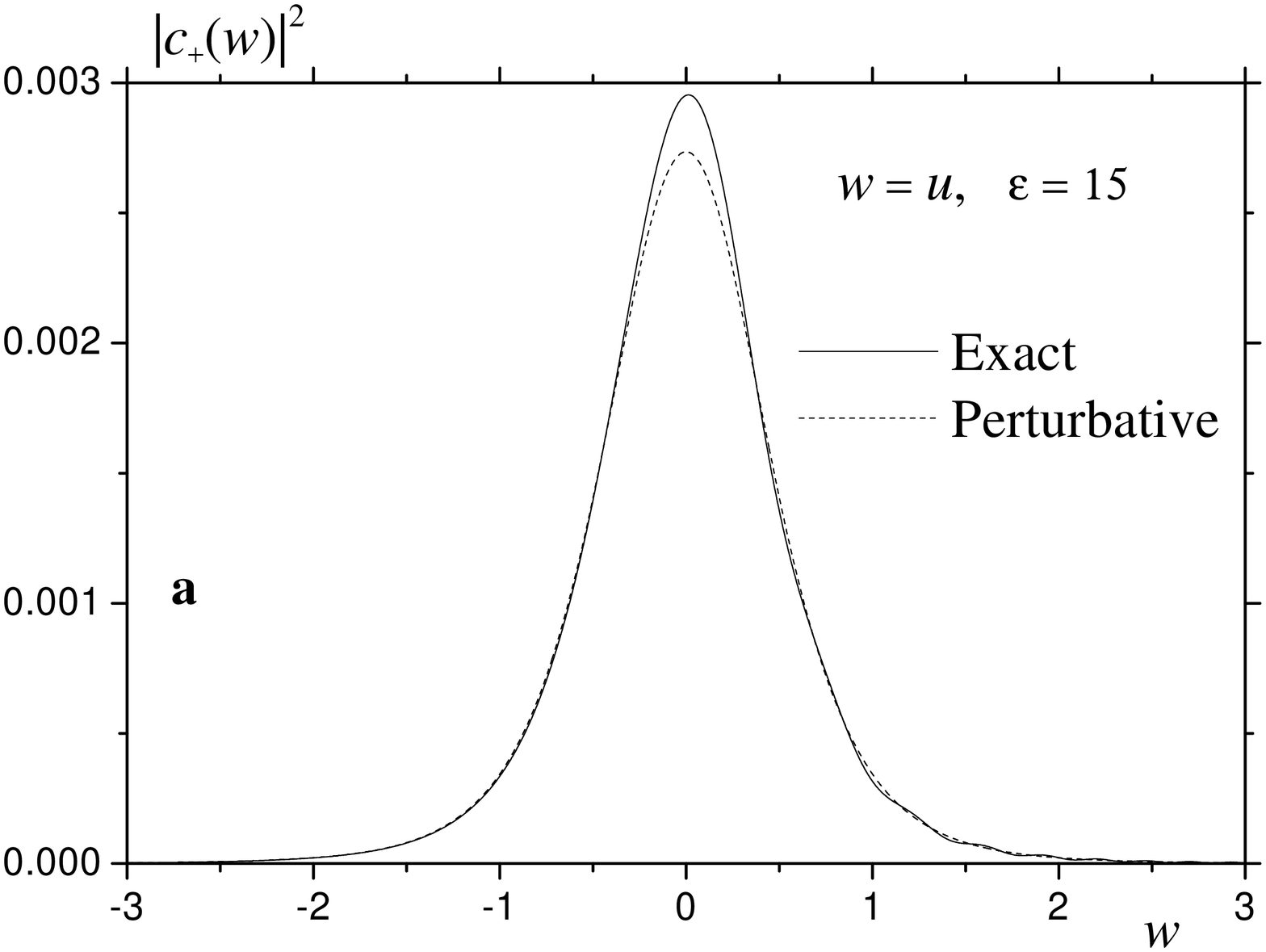,angle=-90,width=9cm}}
\end{picture}
\begin{picture}(11,6)
\centerline{\psfig{file=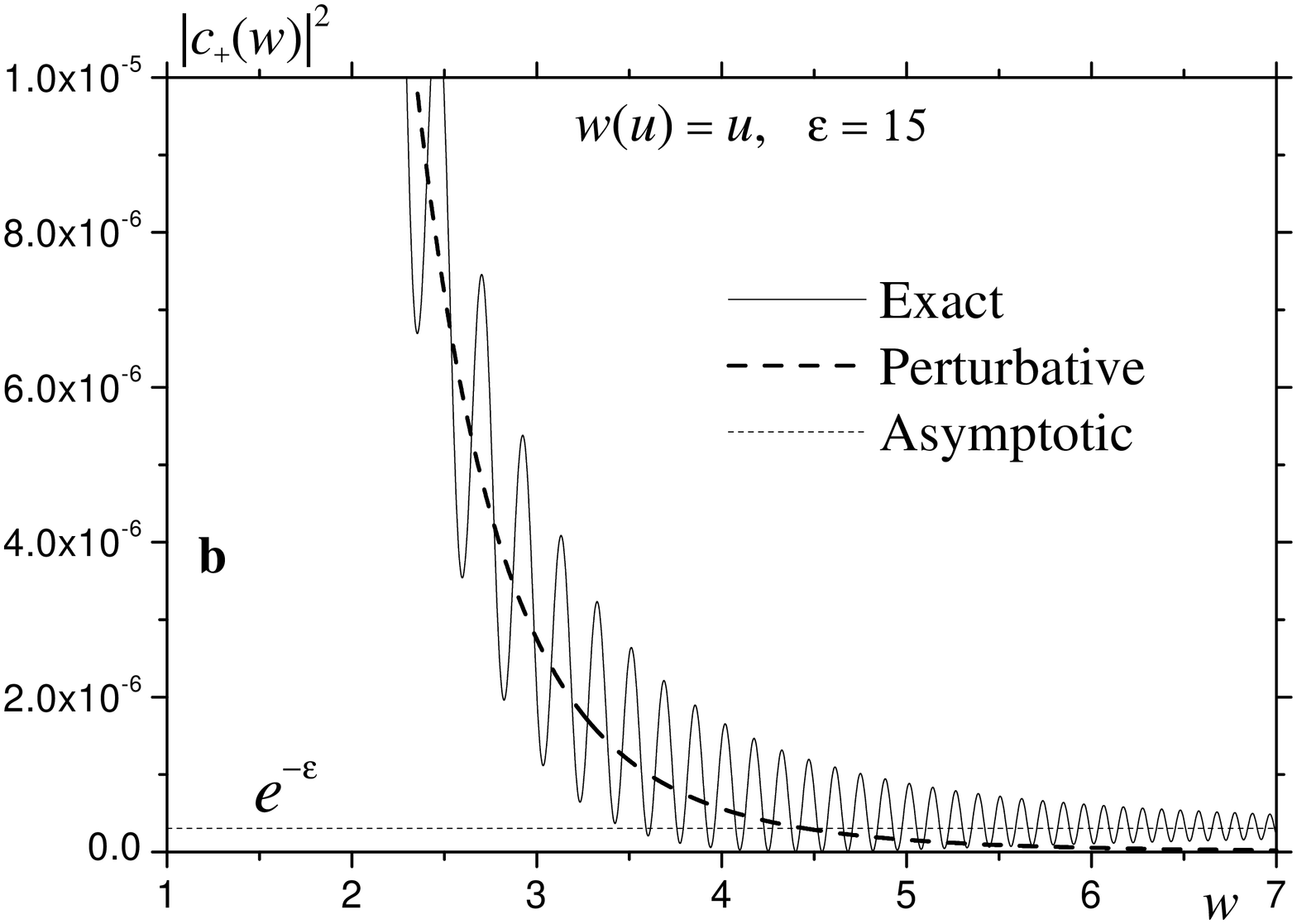,angle=-90,width=9cm}}
\end{picture}
\caption{ \label{fig-cpw}
Lowest-order $1/\tilde\varepsilon$ expansion for $|\tilde{c}_{+}(w)|_{{\rm pert}}^{2}$ compared with the exact numerical solution.
$a$-General view; $b$-Right-wing magnification.
}
\end{figure}
}%

To obtain the solution of the problem that is improved by use of the $1/%
\tilde{\varepsilon}$ expansion while not losing the singular
contribution at $w\rightarrow \infty ,$ one should substitute Eq.\
(\ref{cpmepsExp}) into Eq.\ (\ref{ctilpgenW}). After multiple
integration by parts the latter takes up the form
\begin{equation}
P\cong \left| \frac{1}{2}\int_{-\infty }^{\infty }dw\,e^{i\tilde{\varepsilon}%
\Phi (w)}\sum_{n=0}^{\infty }\left( -u^{\prime }(w)\overline{\Omega }(w)%
\hat{I}\right) ^{n}\frac{\tilde{c}_{-,n}(w)}{\overline{\Omega }^{2}(w)}%
\right| ^{2},  \label{PAdibyParts}
\end{equation}
$\hat{I}$ being the integration operator. The integration by parts
eliminated the powers of $1/\tilde{\varepsilon}$ that are present in Eq.\ (%
\ref{cpmepsExp}). On the other hands, it is clear that contributions of the
terms with $n>0$ in Eq.\ (\ref{PAdibyParts}) should be small for $\tilde{%
\varepsilon}\gg 1$ unless these terms have singularities at $w=i.$
These singularitities do exist, as can be seen from Eq.\
(\ref{Recurrence}). So in the leading order in
$1/\tilde{\varepsilon}$ it is sufficient to take into
account only the most singular contributions in $\tilde{c}_{\pm n}(w)$ near $%
w=i$ that allows to simplify Eqs.\ (\ref{cpcmEqsWRec}) using
\begin{eqnarray}
w &=&i+z,\qquad |z|\ll 1,  \nonumber \\
\qquad \overline{\Omega } &\cong &\sqrt{2iz},\qquad u^{\prime
}=u_{c}^{\prime }={\rm const.}  \label{SimplNeari}
\end{eqnarray}
Then Eqs.\ (\ref{cpcmEqsWRec}) become
\begin{eqnarray}
\frac{d\tilde{c}_{-,n}(z)}{dz} &=&\frac{\tilde{c}_{+,n}(z)}{4iz}  \nonumber
\\
\tilde{c}_{+,n}(z) &=&-\frac{1}{u_{c}^{\prime }\sqrt{2iz}}\left[ \frac{d%
\tilde{c}_{+,n-1}(z)}{dz}+\frac{\tilde{c}_{-,n-1}(z)}{4iz}\right] .
\label{cpcmEqsWRecz}
\end{eqnarray}
The solution of these equations can be searched in the form
\begin{equation}
\tilde{c}_{\pm ,n}(z)=\left\{
\begin{array}{c}
i \\
1
\end{array}
\right\} \frac{\beta _{\pm ,n}}{(2i)^{n/2}\left( u_{c}^{\prime }\right)
^{n}z^{3n/2}},  \label{ctilpmnAnsatz}
\end{equation}
where the coefficients $\beta _{\pm ,n}$ satisfy
\begin{eqnarray}
-6n\beta _{-,n} &=&\beta _{+,n},\qquad \beta _{-0}=1  \nonumber \\
\beta _{+,n} &=&\frac{3}{2}(n-1)\beta _{+,n-1}+\frac{1}{4}\beta _{-,n-1}
\label{betapmnEqs}
\end{eqnarray}
and can be obtained recurrently. Under the same conditions, the sum in Eq.\ (%
\ref{PAdibyParts}) simplifies to
\begin{eqnarray}
&&\sum_{n=0}^{\infty }\left( -u^{\prime }\overline{\Omega }\hat{I}\right)
^{n}\frac{\tilde{c}_{-,n}}{\overline{\Omega }^{2}}  \nonumber \\
&&\qquad {}\cong \sum_{n=0}^{\infty }\left( -u_{c}^{\prime }\sqrt{2iz}\hat{I}%
\right) ^{n}\frac{\beta _{-,n}}{(2i)^{n/2+1}\left( u_{c}^{\prime }\right)
^{n}z^{3n/2+1}}  \nonumber \\
&&\qquad {}=\frac{1}{2iz}\sum_{n=0}^{\infty }\left( \frac{2}{3}\right) ^{n}%
\frac{\beta _{-,n}}{n!}=\frac{1}{2iz}\sum_{n=0}^{\infty }\delta _{n},
\label{SumOvern}
\end{eqnarray}
where we used $\hat{I}z^{a}=z^{a+1}/(a+1)$ and defined $\delta
_{n}=(2/3)^{n}\beta _{-,n}/n!$ For $\delta _{n}$ one obtains from
Eqs.\ (\ref {betapmnEqs}) the recurrence relation
\[
\delta _{n}=\left( 1-\frac{5}{6n}\right) \left( 1-\frac{7}{6n}\right) \delta
_{n-1},\qquad \delta _{0}=1.
\]
Its solution is
\begin{eqnarray*}
\delta _{n} &=&\prod_{k=1}^{n}\left( 1-\frac{5}{6k}\right) \left( 1-\frac{7}{%
6k}\right)  \\
&=&\frac{\left( 1/6\right) _{n}\left( -1/6\right) _{n}}{\left( n!\right) ^{2}%
}=\frac{\left( 1/6\right) _{n}\left( -1/6\right) _{n}}{\left( 1\right) _{n}}%
\frac{1}{n!},
\end{eqnarray*}
where $(a)_{n}\equiv \Gamma (n+a)/\Gamma (a).$ The sum
$\sum_{n=0}^{\infty }\delta _{n}$ in Eq.\ (\ref{SumOvern}) is a
hypergeometric function of argument 1:
\begin{eqnarray*}
\sum_{n=0}^{\infty }\delta _{n} &=&\sum_{n=0}^{\infty }\left. \frac{\left(
1/6\right) _{n}\left( -1/6\right) _{n}}{\left( 1\right) _{n}}\frac{x^{n}}{n!}%
\right| _{x=1} \\
&=&\left. _{2}F_{1}(1/6,-1/6;1;x)\right| _{x=1} \\
&=&\frac{1}{\Gamma (1+1/6)\Gamma (1-1/6)}=\frac{3}{\pi }.
\end{eqnarray*}
Note that the approximation $\tilde{c}_{-}(w)\Rightarrow 1$ in
Eq.\ (\ref {ctilpgenW}) that leads to Eq.\ (\ref{PAdiLin}) for the
prefactor amounts to neglection of all the terms of this sum
except for the zeroth term $\delta _{0}=1.$ Summing up all the
most singular terms of the $1/\tilde{\varepsilon} $ expansion, as
was done above, compensates for the wrong factor $(\pi /3)^{2}$ in
Eq.\ (\ref{PAdiLin}) and renders the prefactor the correct value
1. This value of the prefactor is valid for both linear {\em and}
nonlinear sweep, if $P$ is dominated by only one singularity in
the complex plane. It is the consequence of cancelling of the
quantity $u_{c}^{\prime }$, which describes the nonlinearity of
the sweep, in Eqs.\ (\ref{PAdiLin}) and (\ref {SumOvern}).

We thus have obtained the formula for the probability to stay in the initial
unperturbed state 1 after a slow ($\tilde{\varepsilon}\gg 1$) energy sweep
through the resonance
\begin{equation}
P\cong e^{-\eta },\qquad \eta =2\tilde{\varepsilon}%
\mathop{\rm Im}%
\Phi _{c},\qquad \Phi _{c}=\int^{u_{c}}du\overline{\Omega }(u),
\label{PAdiOneSing}
\end{equation}
where $\tilde{\varepsilon}$ is given by Eq.\ (\ref{epstildeDef}), $\overline{%
\Omega }(u)=\sqrt{1+w^{2}(u)}$ is the dimensionless energy gap at the
avoided level crossing. Note that $\Phi _{c}$ does not depend on $\tilde{%
\varepsilon}.$ The integral in Eq.\ (\ref{PAdiOneSing}) is taken
from the real axis to the singularity point $u_{c}$ in the upper
complex plane that is defined by $\overline{\Omega }(u_c)=0$. This
formula can be found in the textbook by Landau and Lifshitz.
\cite{lanlif3} Pokrovsky, Savvinykh, and Ulinich
\cite{poksavuli58} showed for a similar problem of the overbarrier
reflection of a quantum-mechanical particle that the prefactor is
equal to 1, using a different method. For a weakly-nonlinear sweep
described by Eq.\ (\ref{fDef}) the exponent $\eta $ is
given by Eq.\ (\ref{etaquasilinear}). In contrast, as we have seen
in Sec.\ \ref{sec-nonanalyticslow}, for {\em non-analytic} sweep functions $%
W(t)$ the result is {\em not} exponentially small and is given by
Eq.\ (\ref {PAdiSingRes}) for a particular case of crossing and
returning power-law sweep functions of Eq.\ (\ref{WbarPowDef}).

\subsubsection{A pair of singularities}

In many important cases the behavior of $P$ is more complicated
than the well-known formula Eq.\ (\ref{PAdiOneSing}) does suggest.
For essentially nonlinear time-antisymmetric (crossing) or
symmetric (returning) sweep functions $W(t)$ typically there are
{\em two} singularities at $u_{c\pm }$ in the upper complex plane
that are closest to the real axis and symmetric with respect to
the imaginary axis. It can be shown with the same method that $P$
has the form ($\tilde{\varepsilon}\gg 1$)
\begin{equation}
P\cong \left| e^{i\tilde{\varepsilon}\Phi _{c+}}+e^{i\tilde{\varepsilon}\Phi
_{c-}}\right| ^{2},\qquad \Phi _{c\pm }=\pm \Phi _{c}^{\prime }+i\Phi
_{c}^{\prime \prime },  \label{PAdiTwoPoles}
\end{equation}
i.e., both contributions have prefactors 1. This is not surprising
since for $\tilde{\varepsilon}\gg 1$ contributions from different
singularities are well separated from each other and the method
used above applies to each of them. Eq.\ (\ref{PAdiTwoPoles}) can
be rewritten in the form
\begin{equation}
P\cong P_{0}e^{-\eta },\qquad P_{0}=4\cos ^{2}\left( \tilde{\varepsilon}\Phi
_{c}^{\prime }\right) ,\qquad \eta =2\tilde{\varepsilon}\Phi _{c}^{\prime
\prime }  \label{PAdiTwoPolesRes}
\end{equation}
that contains an oscillating prefactor $P_{0}$ with $\Phi
_{c}^{\prime }\sim $ $1,$ according to the definition of the phase
$\Phi $ in Eq.\ (\ref {ctilpgenW}). These results pertain to the
crossing sweep. For the returning sweep one obtains
\begin{equation}
P\cong 1-P_{0}e^{-\eta }.  \label{PAdiTwoPolesResRet}
\end{equation}

Turning the prefactor $P_{0}$ to zero for some (finite!) sweep
rates is the so-called complete conversion from state 1 to state 2
that was mentioned in the introduction. In Ref.\
\onlinecite{garsch02} we have found this phenomenon for specially
chosen sweep functions $W(t)$ of a more complicated form. Here we
demonstrated that the full conversion at finite sweep rates is a
general phenomenon for a nonlinear sweep. Whereas an oscillating
prefactor can be found in earlier publications (see, e.g., Ref.\
\onlinecite{ternak97}), here we have shown its simple relation to
the relevant singularities of the LZS problem.

\begin{figure}[t]
\unitlength1cm
\begin{picture}(11,5)
\centerline{\psfig{file=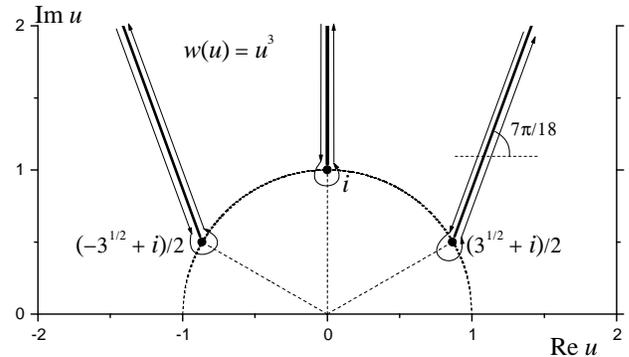,angle=-90,width=9cm}}
\end{picture}
\caption{ \label{Contour} Integration contour in the complex plane
$u$ in Eq.\ (\protect\ref{ctilpgenW}) for the sweep function
$w(u)=u^3$. A pair of singularities at $u=(\pm\sqrt{3}+i)/2$
dominate the probability $P$ for slow sweep, $\varepsilon\gg 1$. }
\end{figure}

\subsection{\protect\bigskip Particular cases}

\begin{figure}[t]
\unitlength1cm
\begin{picture}(11,6)
\centerline{\psfig{file=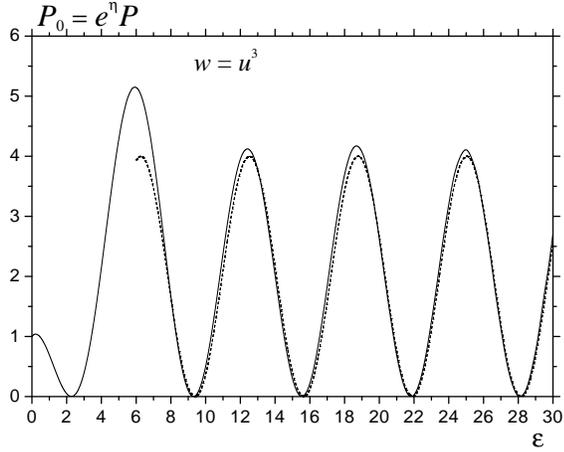,angle=-90,width=9cm}}
\end{picture}
\caption{ \label{P0alpha3} Prefactor $P_0$ in $P(\varepsilon)=P_0
e^{-\eta}$ for power-law sweep function $w=u^3$. The solid line is
$e^{\eta}P$ with numerically computed $P$ and
$\eta=2\tilde\varepsilon\Phi _{c}^{\prime \prime }$ and the dashed
line is the analytical result $P_{0}=4\cos^{2}\left(
\tilde\varepsilon\Phi _{c}^{\prime }\right)$ [see Eqs.\
(\protect\ref{PAdiTwoPolesRes}) and (\protect\ref{PhiPower})].}
\end{figure}

Let us now consider particular cases of the LZS effect with
nonlinear energy sweep, which will allow to explore the role of
the singularities in the prefactor in more detail. For the
power-law sweeps of Eq.\ (\ref{WbarPowDef})
with natural $\alpha=n$, such as $w(u)=u^{2},$ $w(u)=u^{3},$ etc., there are $%
n$ singularities in the upper complex plane at $u=u_{k}=\exp \left[ i\pi
(1/2+k)/n\right] $ with $k=0,1,\ldots ,n-1$ (see Fig.\ \ref{Contour} for $%
\alpha =3$). Closest to the real axis are those with $k=0$ and
$k=n-1$.
The integration contour in Fig.\ \ref{Contour} goes along both
sides of the cuts initiating at the singularity points. These cuts
are chosen so that $\delta\Phi$ in Eq.\ (\ref{SWDef}) is real on
both sides of the cut (i.e., the cuts are the so-called
anti-Stokes lines), so that $P$ is determined by purely
oscillating integrals. For all other directions of the cuts, one
has exponentially converging integrals on one side of the cut but
exponentially diverging integrals on another side of the cut.
After calculation of $\Phi _{c}$ in Eq.\ (\ref{PAdiOneSing}) one
obtains
\begin{equation}
\left\{
\begin{array}{l}
\Phi _{c}^{\prime } \\
\Phi _{c}^{\prime \prime }
\end{array}
\right\} =\left\{
\begin{array}{l}
\cos \frac{\pi }{2n} \\
\sin \frac{\pi }{2n}
\end{array}
\right\} 2^{1/n}\frac{\Gamma ^{2}\left( 1+\frac{1}{2n}\right) }{\Gamma
\left( 2+\frac{1}{n}\right) }.  \label{PhiPower}
\end{equation}
Note that for $n=1$ there is only one pole, and Eq.\ (\ref{PhiPower}) yields $%
\Phi _{c}^{\prime }=0$ and $\eta =2\tilde{\varepsilon}\Phi
_{c}^{\prime \prime }=(\pi /2)\tilde{\varepsilon}=\varepsilon ,$
so that one returns to the LZS result $P=e^{-\varepsilon }.$ In
other cases the prefactor $P_{0}$ like in Eqs.\
(\ref{PAdiTwoPolesRes}) and (\ref{PAdiTwoPolesResRet}) oscillates
and turns to zero at some values of the sweep rate, where
contributions of the two singularities in Eq.\
(\ref{PAdiTwoPoles}) cancel each other. In the latter case, a {\em
complete} Landau-Zener-Stueckelberg
transition is achieved for crossing sweeps. We have shown the prefactor $%
P_{0}$ for $w=u^{3}$ in Fig.\ \ref{P0alpha3}, where the solid line is $%
e^{\eta }P$ with numerically computed $P$ and $\eta =2\tilde{\varepsilon}%
\Phi _{c}^{\prime \prime }$ and the dashed line is the analytical result $%
P_{0}=4\cos ^{2}\left( \tilde{\varepsilon}\Phi _{c}^{\prime
}\right) $. It is seen that Eqs.\ (\ref{PAdiTwoPolesRes}) and
(\ref{PhiPower}) work well for large $\varepsilon $. Plotting
$P_{0}$ reveals oscillations that are not seen at large
$\varepsilon $ in Fig.\ \ref{Ppowallarge}.

\begin{figure}[t]
\unitlength1cm
\begin{picture}(11,6)
\centerline{\psfig{file=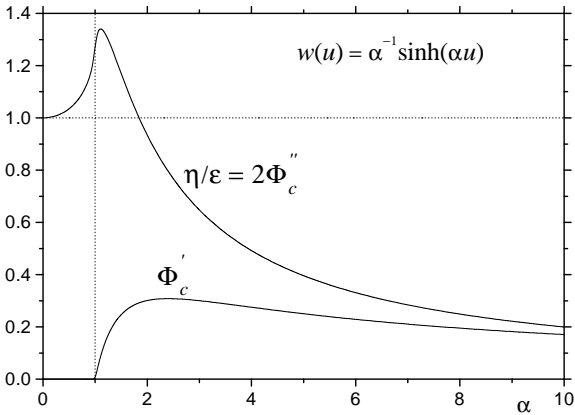,angle=-90,width=9cm}}
\end{picture}
\caption{ \label{Phialphash} Real and imaginary parts of the
quantum-mechanical phase $\Phi$ at symmetric singularities in the
complex plane [see Eq.\ (\protect\ref{PAdiTwoPoles})]  for the
sweep function $w=\alpha^{-1}\sinh(\alpha u)$ vs $\alpha$, given
by Eqs.\ (\protect\ref{etashsmallalp}) and
(\protect\ref{etashlargealp}).}
\end{figure}

\begin{figure}[t]
\unitlength1cm
\begin{picture}(11,6)
\centerline{\psfig{file=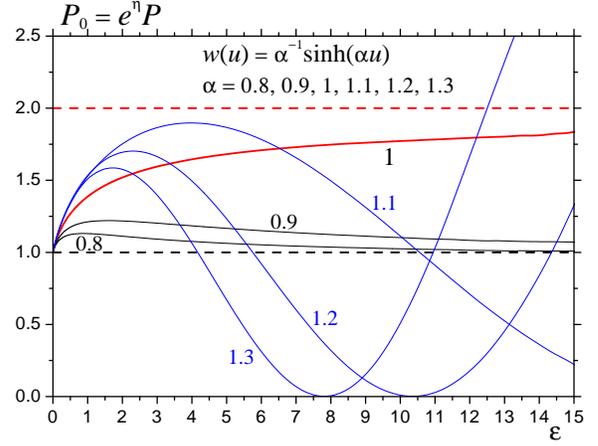,angle=-90,width=9cm}}
\end{picture}
\caption{ \label{P0sharound1} Prefactor $P_0$ in
$P=P_0 e^{-\eta}$ for the sweep function
$w=\alpha^{-1}\sinh(\alpha u)$ defined as $P_0=e^{\eta}P$ with
numerically computed $P$ for $\alpha$ in the vicinity of 1. The dotted lines at $P_0=1$
and at  $P_0=2$ correspond to the analytical results for $\alpha<1$ and $\alpha=1$, respectively.}
\end{figure}

\begin{figure}[t]
\unitlength1cm
\begin{picture}(11,6)
\centerline{\psfig{file=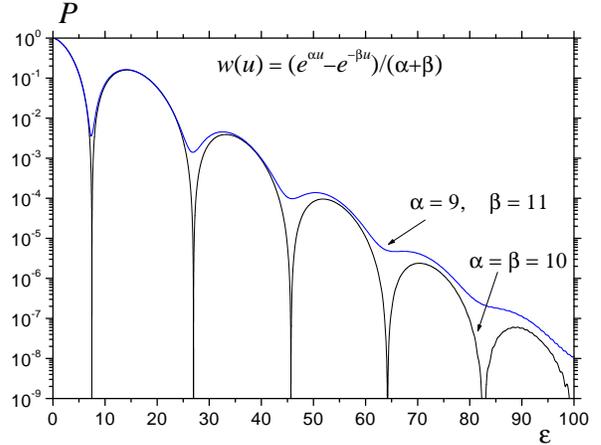,angle=-90,width=9cm}}
\end{picture}
\caption{ \label{shalbet} Staying probability $P$ vs the inverse
sweep rate for the sweep with slightly broken antisymmetry
$w(-u)=-w(u)$. For $\alpha=9$ and $\beta=11$ (same results for
$\alpha=11$ and $\beta=9$) interference of the two poles closest
to the real axis is not complete and $P$ does not turn to zero. }
\end{figure}
%

The next interesting case is that of the sinh sweep function of
Eq.\ (\ref {Wbarsinh}). This model allows to track the changes of
$P$ as function of the continuous parameter $\alpha $ (which is
the measure of nonlinearity)
keeping the sweep function $w(u)$ analytical. The singularities depend on $%
\alpha .$ For $\alpha <1$ the singularity closest to the real axis is at
\begin{equation}
u=u_{c}=i\ \frac{\arcsin \alpha }{\alpha },  \label{ucshsmallalp}
\end{equation}
whereas the next singularity is at $u=i\left( \pi -\arcsin \alpha \right)
/\alpha .$ Keeping only the singularity at $u_{c}$ for $\tilde{\varepsilon}%
\gg 1$ one obtains from Eq.\ (\ref{PAdiOneSing})
\begin{equation}
\eta =\frac{2\tilde{\varepsilon}}{\alpha }{\bf E}\left( \arcsin \alpha ,%
\frac{1}{\alpha ^{2}}\right) ,\qquad (\alpha <1),  \label{etashsmallalp}
\end{equation}
where ${\bf E}(\varphi ,m)$ is the incomplete elliptic integral of second
kind. In the region $\alpha >1$ there is a pair of singularities at
\begin{equation}
u=u_{c\pm }=\pm \frac{1}{\alpha }\ln \left( \sqrt{\alpha ^{2}-1}+\alpha
\right) +\frac{i\pi }{2\alpha }  \label{ucshlargealp}
\end{equation}
that are closest to the real axis. In this case one obtains Eq.\
(\ref {PAdiTwoPolesRes}) with
\begin{eqnarray}
\eta &=&\frac{2\tilde{\varepsilon}}{\alpha }{\bf E}\left( \frac{1}{\alpha
^{2}}\right)  \nonumber \\
\Phi _{c}^{\prime } &=&\frac{1}{\alpha }\left[ {\bf K}\left( 1-\frac{1}{%
\alpha ^{2}}\right) -{\bf E}\left( 1-\frac{1}{\alpha ^{2}}\right) \right] .
\label{etashlargealp}
\end{eqnarray}
Note that $\Phi _{c}^{\prime }=0$ for $0\leq \alpha \leq 1.$ The
$\alpha $ dependences of $\eta $ and $\Phi _{c}^{\prime \prime }$
above are shown in Fig.\ \ref{Phialphash}. In the special case
$\alpha =1$ both singularities join in one and the phase $\Phi
(u)$ of Eq.\ (\ref{SuDef}) becomes a holomorphic function of a
complex argument:
\[
\Phi (u)=w(u).
\]
In this case one could be puzzled by the attempt to apply the
general Landau's arguments \cite{lan32,lanlif3} leading to Eq.\
(\ref{PAdiOneSing}).
Nevertheless, the pole in Eq.\ (\ref{ctilpgen}) due to $1/\overline{\Omega }%
^{2}(u)$ remains and it determines the exponentially small $P.$ It
is again convenient to consider $w$ as the sweep variable and to
use Eq.\ (\ref {ctilpgenW}). The coefficient $\tilde{c}_{-}(w)$
near the pole at $w=i$ is
obtained from Eqs.\ (\ref{cpcmEqsW}) with the Ansatz of Eq.\ (\ref{cpmepsExp}%
), this time taking into account the singularity of
$du/dw=1/\sqrt{1+w^{2}}.$ As the result one arrives at Eq.\
(\ref{PAdiTwoPolesRes}) with the prefactor
\begin{equation}
P_{0}=2,\qquad (\alpha =1).  \label{P0al1}
\end{equation}

Oscillations of the prefactor $P_{0}$ in the region $\alpha >1$
for this model can be illustrated in the same way as was done in
Fig.\ \ref{P0alpha3}. More interesting here is to plot $P_{0}$ in
the vicinity of $\alpha =1$ where the crossover between
oscillating and non-oscillating regimes takes place (see Fig.\
\ref{P0sharound1}). Although the asymptotic numerical behavior of
$P_{0}$ for $\alpha <1$ and $\alpha =1$ at $\varepsilon \gg 1$ is
in accord with the analytical results above, one can see
deviations from this picture at smaller $\varepsilon $ because of
the contribution of the more distant pole at $\alpha <1$.

One can also consider sweep functions that are not time-symmetric
or antisymmetric, such as Eq.\ (\ref{Sweepalphabeta}). If $\alpha
$ and $\beta $ slightly deviate from each other, there are two
singularities that are at slightly different distances from the
real axis. In this case one can observe oscillations of $P$ in
some range of $\varepsilon $
without exactly turning to zero (see Fig.\ \ref{shalbet}). At larger $%
\varepsilon $ the singularity closest to the real axis dominates
and oscillations disappear. Note that the probability $P$ is
symmetric with respect to the interchange of $\alpha $ and $\beta
$ in Eq.\ (\ref {Sweepalphabeta}).

\section{Conclusions and summary}

Our main concern has been the investigation of the LZS effect for nonlinear
sweep. There were three main points we wanted to clarify. {\em First}, what
are the corrections to the LZS propability $P,$ due to weak nonlinearity of
analytical functions $W(t)?$ This question is of primary interest since an
experimental sweep can be linear only approximately. {\em Second}, are there
any qualitatively new features for {\em analytic} but strongly nonlinear $%
W(t)$ and,{\em \ third}, how does $P$ look like for nonanalytical sweep
functions? Of course, the answers to these questions depend on whether the
sweep is fast or slow.

Concerning the first point, we have found for fast sweep rates,
i.e., $\varepsilon \ll 1,$ that $\varepsilon $ sets the scale for
the relevance of nonlinearities of $W(t)$ measured by a
dimensionless parameter $\alpha $ of Eq. (\ref{fDef}). For fast
sweep and weak nonlinearity $\alpha ^{2}\ll \varepsilon \ll 1,$
one obtains the expected result $1-P\cong \varepsilon $ with {\em
positive} corrections of order $\left( \alpha ^{2}/\varepsilon
\right) ^{2}$. For still faster sweeps and/or stronger
nonlinearities, $\varepsilon \ll \min(1,\alpha ^{2})$,  and for
$w(u)$ that is a power series terminating at finite order $u^{n},$
the leading-order result is $1-P\sim \left( \varepsilon /\alpha
^{2}\right) ^{2n/(n+1)}$. This implies that the $\varepsilon
$-dependent transition probability $1-P$ for the fast sweep
changes from $\varepsilon $ for a linear sweep to $\varepsilon
^{2}$ for a nonlinear sweep with $n\rightarrow \infty ,$ in
leading order in $\varepsilon .$ This conclusion, of course, also
follows from the result of Eq.\ (\ref{PPowCrossing}) for the pure
power-law sweep defined by Eq.\ (\ref{WbarPowDef}) since $\alpha $
corresponds to $n.$

The most interesting results follow for {\em slow} sweeps, i.e., $%
\varepsilon \gg 1.$ \ For several qualitatively different sweep functions $%
W(t)$ we have shown that $P$ can exhibit quite different
$\varepsilon$-dependence. For the power-law sweep of Eq.\
(\ref{WbarPowDef}) which is a
nonanalytical function one obtains a power-law behavior for $P,$ i.e., $%
P\sim \varepsilon ^{-2\alpha }$ in case of the crossing sweep. It
would be interesting to experimentally realize such a sweep \
because it should be much easier to measure the $\varepsilon
^{-2\alpha }$ dependence than an exponential dependence
$e^{-\varepsilon },$ for large $\varepsilon .$ For {\em
analytical} sweep functions we have demonstrated that $P$ can
always be decomposed into an exponential factor $e^{-\eta }=\exp
\left( -2\tilde{\varepsilon}\mathop{\rm Im}\Phi _{c}\right) $ and
a prefactor $P_{0}=4\cos ^{2}\left( \tilde{\varepsilon}\mathop{\rm
Re}\Phi _{c}\right) ,$ where $\Phi _{c}$ is determined by the
singularities $u_{c}$ following from
\begin{equation}
w^{2}(u_{c})=-1.  \label{SingCond}
\end{equation}
If $w(u)$ does not deviate strongly from the linear function [see
the exact criteria in the main text] then there is one singularity
in the upper complex plane, only, and one obtains the LZS result
with $P_{0}=1$ and $\eta =\varepsilon .$ However, a new feature
occurs if the energy sweep $W(t)$ exhibits a significant
retardation in the vicinity of the resonance. In this case we have
shown that the prefactor $P_{0}$ becomes oscillatory as function
of $\varepsilon$ and turns to zero at some $\varepsilon$. The
latter corresponds to the complete conversion from state 1 to
state 2. \cite{garsch02} Since the period of the oscillations of
$P_{0}$ is proportional to $\varepsilon ^{-1}\sim v/\Delta ^{2},$
its measurements
would allow independent determination of $\Delta $ for a given sweep rate $%
v, $ besides the measurement of the exponential factor. Of course,
to determine $\Delta $ from the oscillation's period one must
calculate $\mathop{\rm Re}\Phi _{c}$ which is determined by
$W(t)$, only. We mention that oscillations of  $P$ were already
found for different nonlinear-sweep models (see, e.g., Ref.\
\onlinecite{ternak97}), as well as for the exactly solvable
tight-binding electron model on one-dimensional chain driven by an
electric field.\cite{poksin02}

Oscillations of the prefactor $P_{0}$ in Eq.\
(\ref{PAdiTwoPolesRes}) for essentially nonlinear energy sweeps
find physical explanation. Consider, for instance, the sinh sweep
of Eq.\ (\ref{Wbarsinh}). The system is in the vicinity of the
resonance for $-u_{0}\lesssim u\lesssim u_{0},$ where $u_{0}$
satisfies $w(u_{0})=1$ and is given by $u_{0}=\ln \left( \sqrt{\alpha ^{2}+1}%
+\alpha \right) /\alpha .$ For $\alpha \gg 1$ one has $u_{0}\ll 1$ and the
derivative $w^{\prime }(u_{0})=\cosh \left( \alpha u_{0}\right) =\sqrt{%
\alpha ^{2}+1}$ is large, thus $w(u)\ll 1$ in the main part of the interval $%
-u_{0}\lesssim u\lesssim u_{0}.$ This means that for $\alpha \gg
1$ the system rapidly comes into the vicinity of the resonance,
stays practically at resonance for some time, and then rapidly
leaves the resonance region. During this stay, the system can
oscillate between states 1 and 2, so that the value of $P$ depends
on the time spent at resonance and at some values of
$\tilde{\varepsilon}$ it turns to zero. The time spent at resonance $%
t_{0}=\left( \Delta /v\right) u_{0}$ [see Eq.\ (\ref{WbarDef})] is
controlled by the sweep rate $v$ or by the parameter
$\tilde{\varepsilon}=\Delta ^{2}/(\hbar v)$. Note, however, that
for finite $\alpha $ oscillations between states 1 and 2 are not
complete and thus $P<1$ and it becomes exponentially small for
$\tilde{\varepsilon}\gg 1.$ The same physical explanation is valid
for power-law sweeps $w(u)=u^{n}$ with $n\gg 1$, although the
effect is already seen for $n\gtrsim 1$.

Similar situation should be realized in the case of the
overbarrier reflection of a quantum-mechanical particle. If the
potential barrier is parabolic (which corresponds to the linear
sweep in the LZS problem) then the problem has an exact solution
and the reflection coefficient $R$ determined by a single
singularity in the complex plane is a monotonic function of the
particle's energy $E$. \cite{lanlif3} If, however, the potential
$U(x)$ has a flat top, say $U(x)=-x^{2n}$ with $n=2,3,\ldots ,$
then there is a pair of singularities nearest to the real axis that causes $%
R(E)$ to oscillate and turn to zero for some values of $E$. This behavior is
well-known for the rectangular potential that is the limiting case of $%
U(x)=-x^{2n}$ with $n\rightarrow \infty .$

Finally,  the $1/\varepsilon $ expansion of Eq.\ (\ref{cpmepsExp})
that was applied to obtain the prefactor in the LZS formula for
analytical sweep functions and slow sweeps might be interesting
from the technical point of view.

\section*{Acknowledgments}

We would like to thank Eugene Chudnovsky for useful discussions.

\bibliographystyle{prsty}

\begin{thebibliography}{10}
\bibitem{LZFoot}
Landau \cite{lan32} obtained this formula without the prefactor in
the adiabatic case $\varepsilon \ll 1$; Zener's exact result
\cite{zen32} is wrong by a factor $2\pi $ in the exponential;
Akulin and Schleich\cite{akusch92} included decay from state 2
that, in a remarcable way, does not change the result.
\end{thebibliography}

\end{document}